\documentclass[aps,prd,twocolumn,superscriptaddress,nofootinbib,floatfix,noshowpacs]{revtex4-2}
\pdfoutput=1
\newif\ifcomment
\usepackage{amsmath,amssymb}
\usepackage{color,hyperref,url}
\usepackage{listings}
\usepackage{slashed}
\usepackage[pdftex]{graphicx}
\usepackage{epstopdf}
\usepackage{epsfig}
\usepackage{grffile}
\usepackage{relsize}
\graphicspath{{./img/}}
\usepackage{soul}

\newcommand{\beq}{\begin{equation}}
\newcommand{\eeq}{\end{equation}}
\newcommand{\ba}{\begin{array}}
\newcommand{\ea}{\end{array}}
\newcommand{\bea}{\begin{align}}
\newcommand{\eea}{\end{align}}
\newcommand{\bi}{\begin{itemize}}
\newcommand{\ei}{\end{itemize}}
\newcommand{\ben}{\begin{enumerate}}
\newcommand{\een}{\end{enumerate}}
\newcommand{\bc}{\begin{center}}
\newcommand{\ec}{\end{center}}
\newcommand{\bl}{\begin{flushleft}}
\newcommand{\el}{\end{flushleft}}
\newcommand{\br}{\begin{flushright}}
\newcommand{\er}{\end{flushright}}

\renewcommand{\>}{\rangle}   

\begin{document}
\title{Pion scalar, vector and tensor form factors from a contact interaction}
\author{Xiaobin Wang}%
\affiliation{School of Physics, Nankai University, Tianjin 300071, China}
\author{Zanbin Xing}%
\affiliation{School of Physics, Nankai University, Tianjin 300071, China}
\author{Jiayin Kang}%
\affiliation{School of Physics, Nankai University, Tianjin 300071, China}
\author{Kh\'epani Raya}
\email{khepani.raya@dci.uhu.es}
\affiliation{Department of Integrated Sciences and Center for Advanced Studies in Physics, Mathematics and Computation, University of Huelva, E-21071 Huelva, Spain.}
\author{Lei Chang}%
\email{leichang@nankai.edu.cn}
\affiliation{School of Physics, Nankai University, Tianjin 300071, China}
\date{\today}
\begin{abstract}
The pion scalar, vector and tensor form factors are calculated within a symmetry-preserving contact interaction model (CI) of quantum chromodynamics (QCD), encompassed within a Dyson-Schwinger and Bethe-Salpeter equations approach. In addition to the traditional rainbow-ladder truncation, a modified interaction kernel for the Bethe-Salpeter equation is adopted. The implemented kernel preserves the vector and axial-vector Ward-Takahashi identities, while also providing additional freedom. Consequently, new tensor structures are generated in the corresponding interaction vertices, shifting the location of the mass poles appearing in the quark-photon and quark tensor vertex and yielding a notorious improvement in the final results. Despite the simplicity of the CI, the computed form factors and radii are compatible with recent lattice QCD simulations.
\end{abstract}
\keywords{...}

\maketitle
\section{introduction}\label{sec:int}
Quantum Chromodynamics (QCD) is regarded as the underlying theory that describes the strong interactions in the Standard Model. This quantum field theory is full of complexities that arise due to its non-Abelian nature, producing very peculiar features in the characterizing running coupling~\cite{Cui:2019dwv}, thereby limiting the usage of perturbation theory to certain kinematic regimes~\cite{Lepage:1980fj,Efremov:1979qk}. Giving this notorious shortcoming, to address several hadron properties and low-energy phenomena, such as confinement and emergent hadronic mass (EHM)~\cite{Roberts:2020hiw,Roberts:2021nhw}, non-perturbative frameworks turn out to be necessary. For the past few decades, the Dyson-Schwinger Equations formalism has proven to be a powerful tool to deal with QCD in the continuum~\cite{Roberts:1994dr,Fischer:2018sdj}, paving a traceable connection between the structural properties of hadrons and the fundamental degrees of freedom of QCD, quarks and gluons. Being both a QCD boundstate and a (pseudo) Nambu-Goldstone boson (NGb), the pion takes center stage in our understanding of the strong interactions, specially concerning the non-pertubative phenomena in QCD and EHM~\cite{Horn:2016rip,Arrington:2021biu}. Herein, the pion structure is investigated via the computation of form factors (FFs). The pion vector form factor, namely, the electromagnetic form factor (eFF), describes the coupling of a photon to the pion. It is then directly accessible from experiments~\cite{NA7:1986vav,Bebek:1977pe,Barkov:1985ac,Bebek:1974ww,Quenzer:1978qt,DM2:1988xqd,Amendolia:1983di,CMD-2:2005mvb,CMD-2:2006gxt,JeffersonLabFpi:2000nlc,JeffersonLabFpi-2:2006ysh,JeffersonLab:2008jve} and invesgated through multiple approaches, such as lattice QCD simulations~\cite{Meyer:2011um,Frezzotti:2008dr,Boyle:2008yd,Feng:2014gba,Gao:2021xsm}, Continuum Schwinger Methods (CSM)~\cite{Miramontes:2021exi,Miramontes:2021xgn,Eichmann:2019tjk,Chen:2018rwz,Chang:2013nia,Maris:2000sk,Maris:1999bh} and many other phenomenological approaches~\cite{Guerrero:1997ku,Bakulev:2004cu,Brodsky:2007hb,Ghorbani:2011gc,Djukanovic:2014rua,Ananthanarayan:2020vum,Cheng:2020vwr,MartinContreras:2021yfz}. Conversely, the pion scalar and tensor form factors, which are not  directly accessible as such from experiments, have not received sufficient attention. Notwithstanding, these two form factors still deserve scrutiny. For instance, the scalar form factor (sFF) can be connected with the $\pi\pi$ scattering processes, of extreme relevance in nuclear physics, and can be used to analyze the potentially existing scalar mesons~\cite{Dubnicka:2016bhn}; on the other hand, the calculation of the tensor form factor (tFF) is complementary to explorations concerning gravitational FFs~\cite{Freese:2019bhb,Zhang:2020ecj}, and could shed some light in our understading of the interaction of the pion with higher-spin probes, as it might happen in beyond Standard Model theories. For comprehensive reviews on the pion scalar and tensor FFs, see Refs.~\cite{Alexandrou:2021ztx,Gulpers:2013uca,JLQCD:2008kdb,Caprini:2018jjm,Oller:2007xd,Hannah:1999ev,JLQCD:2009ofg,QCDSF:2007ifr}. 

In the present work, we calculate the pion scalar, vector and tensor FFs within the DSE approach. Based upon the traditional rainbow-ladder (RL) truncation in  QCD~\cite{Bender:1996bb,Munczek:1994zz}, a modified rainbow-ladder (MRL) truncation is adopted. As detailed in Ref.~\cite{Xing:2021dwe}, both truncations preserve the vector and axial-vector Ward-Green-Takahashi identities (WGTIs), ensuring charge conservation and NGb nature of the pion, while also introducing timelike mass poles in the relevant interaction vertices (\emph{e.g.} a vector meson pole in the quark-photon vertex). The additional structures present in the MRL truncation generate a quark anomalous magnetic moment (AMM) term, whose impact could be noticeable in meson and baryon FFs~\cite{Raya:2021pyr,Wilson:2011aa}. The appearance of this term is then desirable and understood as follows: the anomalous chromomagnetic moment for dressed light-quarks, generated non-perturbatively, commensurates (albeit with opposite sign) with its electromagnetic analogue~\cite{Chang:2010hb,Bashir:2011dp}. For illustrative purposes, the calculations presented herein are restricted to the symmetry-preserving contact interaction (CI) model introduced in~\cite{Roberts:2010rn,Gutierrez-Guerrero:2010waf}. We contrast the results produced in the RL and MRL truncations, observing that, despite the simplicity of the CI, those produced in the MRL case are quite compatible with recent lattice QCD simulations~\cite{Alexandrou:2021ztx}.

The manuscript is organized as follows: Section~\ref{sec:ci} introduces the CI within the MRL truncation. Section~\ref{sec:qsvt} focuses on the structure and calculation of the relevant interaction vertices: quark-scalar, vector and tensor vertices. Section~\ref{sec:ff} presents the definition and derivation of the pion scalar, vector and tensor FFs. A summary is the presented in Section~\ref{sec:sum}.

\section{contact interaction model}\label{sec:ci}
Let us now describe the CI model applied to the MRL truncation from Ref.~\cite{Xing:2021dwe} (CI-MRL), which extends the RL case~\cite{Roberts:2010rn} (CI-RL). The DSE for the $f$-flavored quark propagator, often dubbed as gap equation, reads in the CI model~\cite{Roberts:2010rn}:
\begin{equation}
\label{eq:GapEq1}
    S^{-1}_f(p)=[i \gamma \cdot p + m_f]+\frac{4}{3 m_G^2} \int_{q}  \gamma_\mu S_f(q) \gamma_\mu\;,
\end{equation}
where $m_f$ is the Lagrangian current quark mass and $m_{G} = 0.127$ GeV is an infrarred mass scale, the symbol $\int_{q}:= \int d^4q/(2\pi)^4$ denotes a four dimensional Euclidean integral. Note that Eq.~\eqref{eq:GapEq1} posseses a quadratic divergence, such that the integral must be regularized in a Poincaré covariant manner. This implies, firstly, that a general solution implies a quark propagator written as
\begin{equation}
S^{-1}_f(p)=i\gamma \cdot p +M_f,
\end{equation}
where $M_f$ is a momentum independent mass function. The gap equation thus becomes
\begin{equation}
M_f=m_f+\frac{M_f}{3\pi^2 m_{G}^{2}}\int_{0}^{\infty}ds\frac{s}{s+M_f^2}\,.
\end{equation}
Using proper time regularization, the integral above is recast as:
\begin{equation}
\frac{1}{s+M_f^2}=\int_{0}^{\infty}d\tau e^{-\tau (s+M_f^2)}\rightarrow\int_{\tau_{uv}^2}^{\tau_{ir}^2}d\tau e^{-\tau (s+M_f^2)}\,,
\end{equation}
where $\tau_{uv}$ and $\tau_{ir}$ are infrared and ultraviolet regulators, respectively. A nonzero value of $\tau_{ir}:= 1/\Lambda_{ir} =1/0.24$ GeV$^{-1}$  ensures the absence of quark production thresholds, thereby being compatible with confinement. On the other hand, since the contact interaction does not define a renormalizable theory, $\tau_{uv}:= 1/\Lambda_{uv}=1/0.905$ GeV$^{-1}$ cannot be removed and instead plays a dynamical role, setting the scale of all dimensioned quantities. With the regularization prescription we have introduced, the quark mass function is obtained from:
\begin{equation}
M_f=m_f+\frac{M_f}{3\pi^2 m_{G}^{2}}\mathcal{C}^{iu}(M_f^2),
\end{equation}
where $\mathcal{C}^{iu}(z)/z=\Gamma(-1,z\,\tau_{uv}^2)-\Gamma(-1,\,z\tau_{ir}^2)\,,
$ with $\Gamma(a,z)$ the incomplete Gamma function.

The corresponding meson Bethe-salpeter equation (BSE) reads: 
\begin{eqnarray}
\label{eq:BSE1}
\Gamma_H(P)=-\frac{4}{3m_{G}^{2}}\int_{q}\left[\gamma_{\mu}\chi_H(P)\gamma_{\mu} - \xi \tilde{\Gamma}_{j}\chi_H(P)\tilde{\Gamma}_{j}\right],
\end{eqnarray}
where $\chi_H(P) = S_f(q) \Gamma_H(P) S_{\bar{h}}(q-P)$, with $\Gamma_H(P)$ being the Bethe-Salpeter amplitude (BSA); $H$ labels the type of meson (encoding Lorentz indices, if any), and $P$ represents the meson total momentum, such that $P^2=-m_H^2$ ($m_H$ is the meson mass). Note that, setting $\xi=0$ in Eq.~\ref{eq:BSE1}, one recovers the CI-RL case. In this way, we will refer as non-ladder structures to those that are accompanied by $\xi$, and those are: $\tilde{\Gamma}_{n}=\{\mathbb{I},\gamma_5,\frac{i}{\sqrt{6}}\sigma_{\mu\nu}\}$.

Despite the simplicity of CI, significant features of QCD such as confinement and DCSB are preserved within this framework. Furthermore, the CI is proven to be a reliable tool to the meson and baryon spectra, as well as the electro-weak elastic and transition FFs~\cite{Gutierrez-Guerrero:2010waf,Roberts:2010rn,Roberts:2011wy,Roberts:2011cf,Wilson:2011aa,Chen:2012qr,Raya:2021pyr,Segovia:2013uga,Serna:2017nlr,Gutierrez-Guerrero:2021rsx,Xing:2021dwe,Hernandez-Pinto:2022lsi,Xing:2022sor}. From this point on, we will focus on the properties of the pion. The isospin symmetric limit $m_u = m_d$ would be considered, therefore flavor indices shall be omitted for simplicity. The nature of the CI model entails the pion BSA only depends on the total momentum $P$, such that, a general for of the pion BSA reads
\begin{equation}\label{bsa}
\Gamma_{\pi}(P)=\gamma_5\left[ i\,E_\pi(P)+\frac{\gamma \cdot P}{M} F_\pi(P)\right]\;.
\end{equation}
As explained in Ref.~\cite{Xing:2021dwe}, the NL terms in Eq.~\eqref{eq:BSE1} can be recast under Fierz transformation as  $\frac{1}{3}\sigma_{\alpha\beta}\text{tr}_{D}[\sigma_{\alpha\beta}\chi_{H}(P)]$, making apparent that the NL pieces do not contribute in the pseudoscalar and axial-vector channels. Consequently, in the case of the pion, the solutions to the Eq.~\eqref{eq:BSE1} will be the same in both the CI-RL and CI-MRL cases. However, scalar and vector channels are affected by the NL terms; this would naturally impact the corresponding interaction vertices, as discussed below. In fact, a first effect is observed in the structure acquired by the BSA of the $\rho$ meson~\cite{Xing:2021dwe} which, in the CI-MRL case reas
\begin{equation}
    \label{eq:BSArho}
    \Gamma^{\rho}_{\mu}(P)=\gamma_{\mu}^{T}(P)E_{\rho}(P)+\frac{1}{M}\sigma_{\mu\nu}P_{\nu}F_{\rho}(P)\;,
\end{equation}
while in the CI-RL truncation one finds $F_\rho(P) = 0$, so it would be completely defined by the structure 
\begin{equation}
    \gamma_{\mu}^{T}(P)=\gamma_{\mu}-\frac{ \gamma \cdot P \;
  }{P^{2}} \, P_{\mu}\;.
\end{equation}
With these observations in mind, we fix the model parameters as follows: those already appearing in the CI-RL are fixed in order to reproduce the pion mass and decay constant; on the other hand, the strength of the NL pieces, $\xi$, will be set to reproduce the mass of the $\rho$ meson. Model inputs, meson masses and other static properties are collected in  Table~\ref{tab:masscball}. In the next section, we discuss about the structure interaction vertices within this approach: quark-scalar (QS), quark-vector (QV) and quark-tensor (QT) vertices.
\begin{table}[h!]
\caption{\label{tab:masscball} The model inputs:  $m_{G}=0.127$ GeV, $\tau_{uv}=1/0.905$ GeV$^{-1}$, $\tau_{ir}=1/0.24$ GeV$^{-1}$ and $\xi=0.6$. The produced masses and pion properties are listed below. The superscript $c$ denotes the BSAs have been canonically normalized~\cite{Roberts:2010rn}. Mass units in GeV.}
\begin{tabular}{c|cccccc|cc}
\hline
 &$M$ &$m_{\pi}$  & $f_{\pi}$ &$m_{\sigma}$ &$m_{a1}$ &$m_{\rho}$ &$E_{\pi}^{c}$ & $F_{\pi}^{c}$\\
\hline
$m=0.007$ &0.405 & 0.139 & 0.103&0.815 &  1.131 & 0.771 &3.90 &  0.575 \\
\hline
\end{tabular}
\end{table}

\newpage
\section{quark scalar, photon and tensor vertices}\label{sec:qsvt}
\begin{figure}[ht!]
\includegraphics[width=8.6cm]{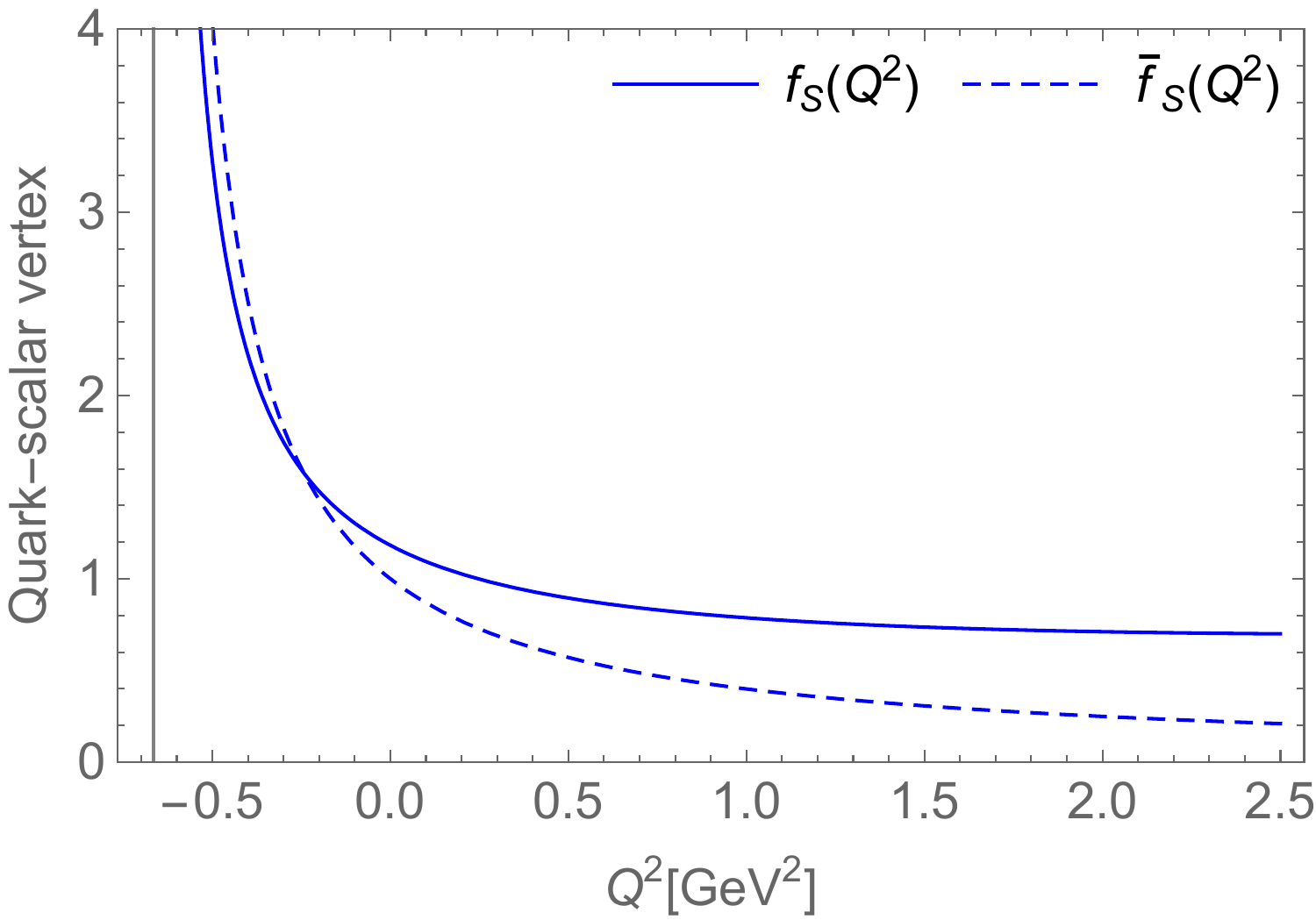}
\caption{Quark-scalar vertex dressing function. The solid line is obtained by solving the quark-scalar vertex BSE, Eq.~\eqref{qsvtbse}, while the dashed line corresponds to the monopole Ansatz from Eq.~\eqref{fsb}. The vertical line in the time-like region indicates the position of the scalar mass pole, \emph{i.e.} $Q^2 = -m_\sigma^2$.\label{qsv}}
\end{figure}

In principle, a fully covariant description of the interaction vertices (herein QS, QP and QT vertices) might require the latter to be characterized by several tensor structures (\emph{e.g.} Refs.~\cite{Miramontes:2019mco, Albino:2018ncl, Maris:1999bh,Krassnigg:2010mh}). In the CI, the seemingly overwhelming task of sensibly determining each set of structures greatly simplifies due to the momentum independent nature of the model; \emph{i.e.} the dressing functions accompanying the different tensor structures do not depend on the relative moment and, therefore, many tensor structures are cancelled by the requirement that its corresponding dressing function be zero. In this way, the QS, QP and QT vertices can be represented as follows:
\begin{eqnarray}
&&\Gamma^{S}(Q)=f_{S}(Q^2)\mathbb{I},\nonumber\\
&&\Gamma_{\mu}^{V}(Q)=f_{V1}(Q^2)\gamma_\mu^L+f_{V2}(Q^2)\gamma_\mu^T+f_{V3}(Q^2)\frac{\sigma_{\mu\nu}Q_\nu}{M},\nonumber\\
&&\Gamma_{\mu\nu}^T(Q)=f_{T1}(Q^2)\sigma_{\mu\nu}+f_{T2}(Q^2)\frac{i}{M}(\slashed{Q}\sigma_{\mu\nu}-\sigma_{\mu\nu}\slashed{Q})\nonumber\\
&&\qquad\qquad+f_{T3}(Q^2)\frac{i^2}{M^2}\slashed{Q}\sigma_{\mu\nu}\slashed{Q}.\label{qsvt}
\end{eqnarray}
where $\gamma_\mu^T=\gamma_\mu-\frac{\slashed{Q}Q_\mu}{Q^2}$, $\gamma_\mu^L=\gamma_\mu-\gamma_\mu^T$; $\Gamma^{\#}(Q)=\{\Gamma^{S}(Q),\ \Gamma_{\mu}^{V}(Q),\ \Gamma_{\mu\nu}^T(Q)\}$ denotes the fully dressed QS, QP and QT vertices, respectively. The vertices satisfy an inhomogeneous BSE, namely:
\begin{eqnarray}
\Gamma^{\#}(Q)&&=\gamma^{\#}-\frac{4}{3m_G^2}\int_q \gamma_\alpha S(q)\Gamma^{\#}(Q)S(q-Q)\gamma_\alpha\nonumber\\
&&+\frac{4\xi}{3m_G^2}\int_q \tilde{\Gamma}_n S(q)\Gamma^{\#}(Q)S(q-Q)\tilde{\Gamma}_n\;.\label{qsvtbse}
\end{eqnarray}
where the inhomogeneous term $\gamma^{\#}=\{\mathbb{I},\ \gamma_{\mu},\ \sigma_{\mu\nu}\}$ clearly depends on the vertex we are describing. To solve this equation, one inserts Eq.~$(\ref{qsvt})$ into Eq.~$(\ref{qsvtbse})$, projects out the different elements on the basis and takes the corresponding Dirac and color traces. This procedure yields a collection of coupled integral equations for the dressing functions $\{f_{S}(Q^2),\ f_{Vi}(Q^2),\ f_{Ti}(Q^2)\}$, whose solutions are plotted through Figs.~(\ref{qsv}-\ref{qtv}). Some details concerning the QT vertex and the subsequent evaluation of the tFF are found in Appendix A (the steps therein detailed are quite general an can be applied to the rest of the cases).

In solving Eq.~$(\ref{qsvtbse})$ for the quark-scalar vertex, one realizes that the profile acquired by $f_S(Q^2)$ turns out to be harder than expected and thus inadequate for a proper description of the sFF. The reason lies within the simple tensor structure acquired by the scalar vertex in the CI framework (a simple identity matrix, Eq.~\eqref{qsvt}), which prevails in both CI-RL and CI-MRL cases. Thus, when calculating the pion scalar form factor, we use the following monopole Ansatz instead:
\begin{equation}\label{fsb}
\bar{f}_{S}(Q^2)=\frac{1}{1+Q^2/m_{\sigma}^2}\,,
\end{equation}
where $m_{\sigma}=0.815$ GeV is the scalar meson mass computed in the CI-MRL scheme, and which also appears as a timelike mass pole in the actual solution of $f_{S}(Q^2)$. Fig.~\ref{qsv} shows the comparisson between the computed QS dressing function $f_{S}(Q^2)$ and the monopole Ansatz, $\bar{f}_{S}(Q^2)$, from Eq.~\eqref{fsb}.

The obtained dressing functions of the quark-photon vertex are depicted in Fig.~\ref{qpv}. With or without NL pieces appearing in the corresponding Bethe-Salpeter kernel, a trivial solution is found for the longitudinal piece of the vertex, \emph{i.e.} $f_{V1}(Q^2)=1$. Appearing in both truncations, $f_{V2}(Q^2)$ features a vector meson pole at $Q^2=-m_\rho^2$, where the $\rho$ meson mass acquires its physical value ($m_\rho = 0.771$  GeV) in the CI-MRL truncation, as opposed to the CI-RL case in which an inflated mass is obtained instead ($m_\rho = 0.929$  GeV in Ref.~\cite{Roberts:2011wy} and $m_\rho = 0.953$ GeV with our preferred value of $m_G$). In addition, it is seen that $f_{V2}(Q^2\to \infty)\to 1$, a contidion that ensures that the the tree level vertex, $\gamma_\mu$, is faithfully recovered in the large $Q^2$ limit. The third dressing function, $f_{V3}(Q^2)$, characterizes the $Q^2$ evolution of the AMM term which, being non-zero only in the CI-MRL case~\footnote{In some occasions, the AMM term is added by hand in CI-RL related calculations~\cite{Wilson:2011aa,Raya:2021pyr}, adopting a particular the form for $f_{V3}(Q^2)$.}, as a consequence of the richer structure of the Bethe-Salpeter kernel, Eq.~\eqref{qsvtbse}. 

Similarly, for the QT vertex, the CI-MRL truncation also produces 3 independent structures, Eq.~\eqref{qsvt}. The corresponding dressing functions are depicted in Fig.~\ref{qtv}. Notably, all of them have a pole at $Q^2=-m_T^2$, where $m_T$ can be regarded as the mass of some intermediate tensor resonance; within our framework, it happens to coincide with $m_\rho$. The existence of the NL term in the Bethe-Salpeter kernel decreases the value $m_T$ so that, as in the QP vertex, the position of the pole move closer to $Q^2=0$. This means a greater influence of the mass pole in the low $Q^2$ domain of FFs, which translates into larger charge radii and a softer behavior in the vicinity of $Q^2=0$. Taking $\xi=0$, \emph{i.e.} the CI-RL case, the quark-tensor vertex dramatically simplifies. There is no longer a pole in the dressing functions $f_{T1,T3}$ and, in fact, those acquire trivial profiles: $f_{T1}=1$ and $f_{T3}=0$. 

\begin{figure}[ht!]
\includegraphics[width=8.6cm]{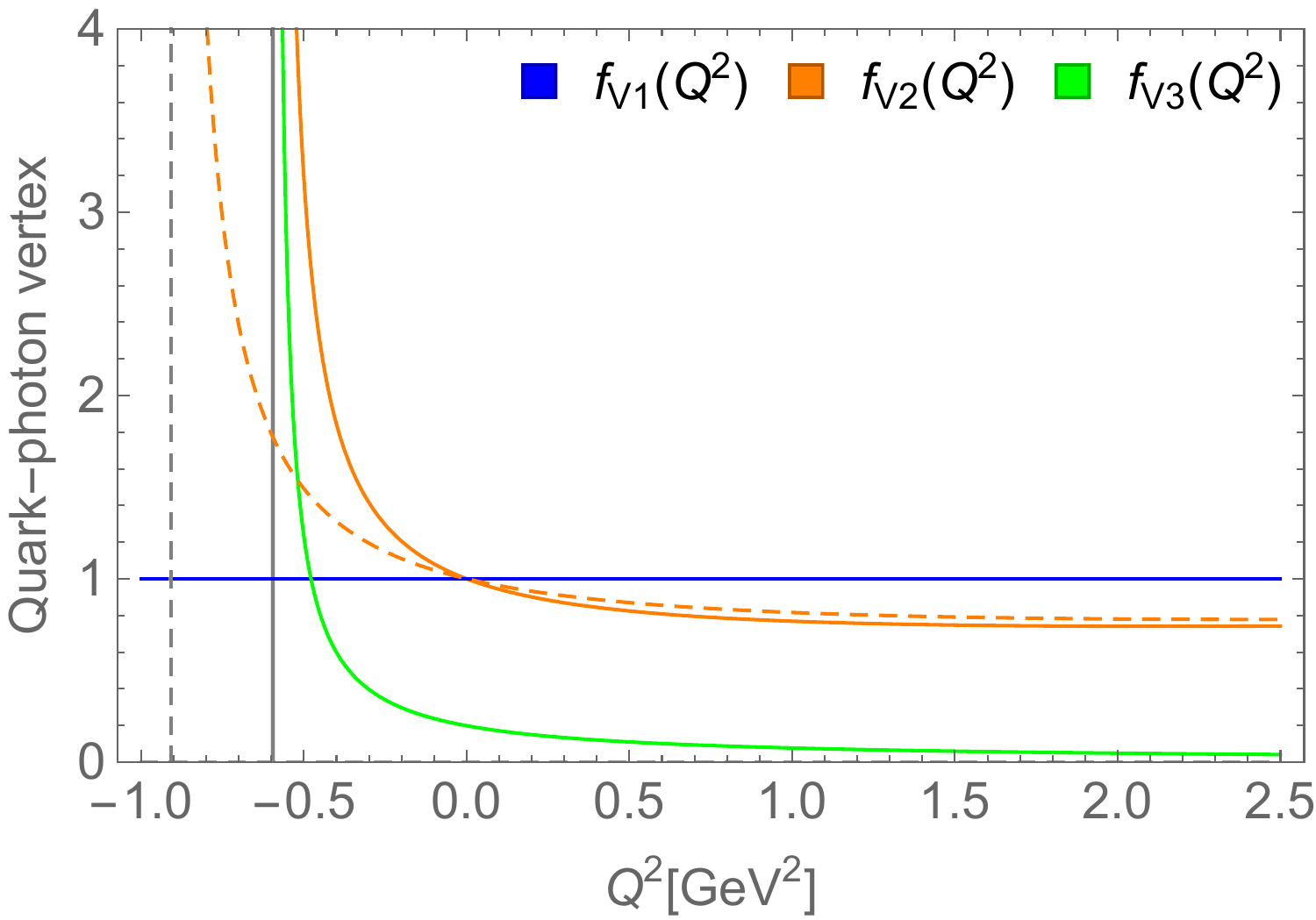}
\caption{Quark-photon vertex dressing functions. The blue, orange and green line stands for 3 dressing functions of the quark-photon vertex defined in Eq.~(\ref{qsvt}). Solid lines corresponds to the CI-MRL case while the dashed lines are obtained in the CI-RL truncation. For each case, the vertical lines in the time-like region denote the indicate of the mass pole of the dressing function $f_{V2}(Q^2)$, \emph{i.e.} $Q^2=-m_\rho^2$.\label{qpv}}
\end{figure}
\begin{figure}[ht!]
\includegraphics[width=8.6cm]{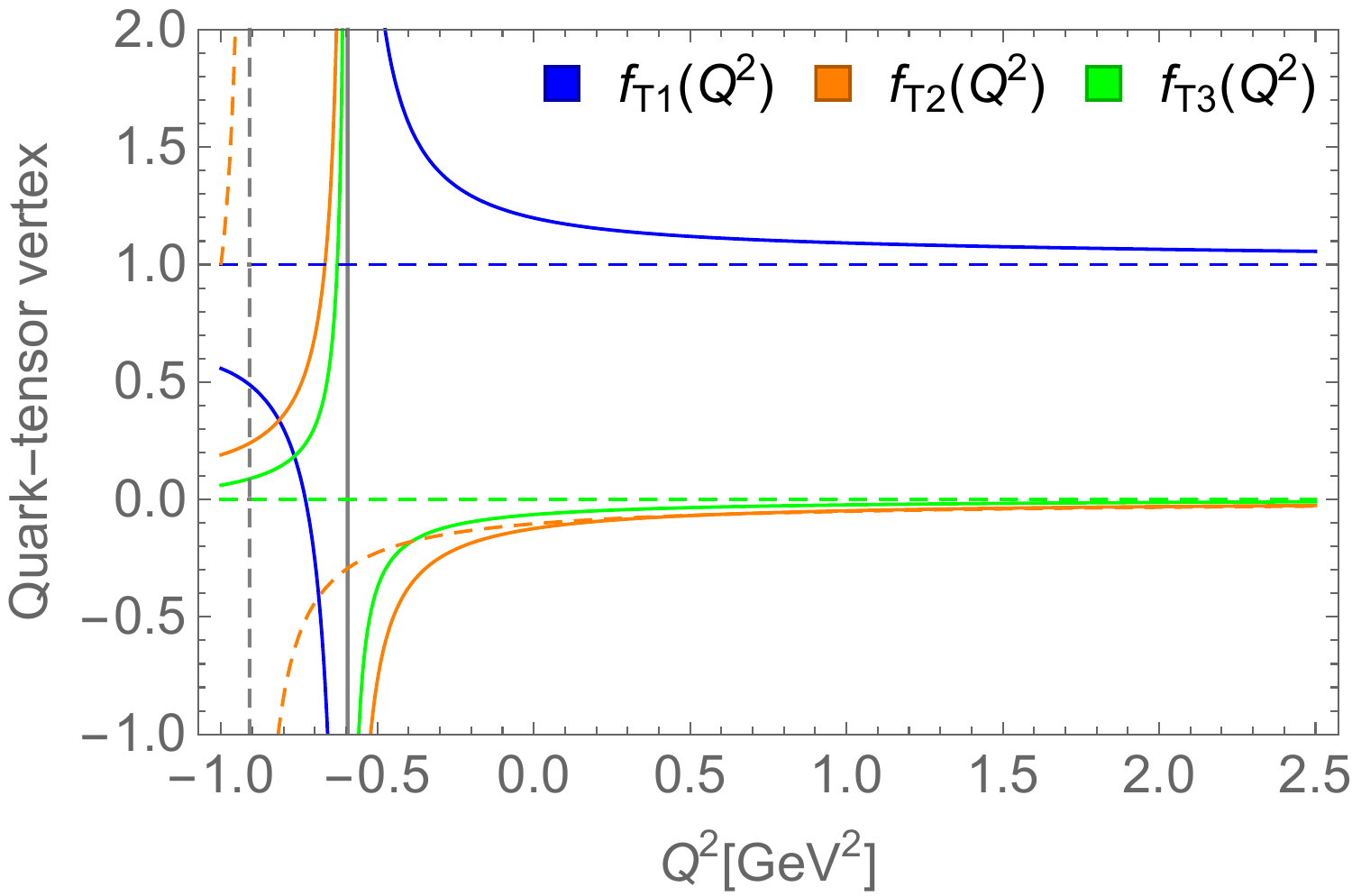}
\caption{Quark-tensor vertex dressing functions. The blue, orange and green line stands for 3 dressing functions of the quark-tensor vertex defined in Eq.~(\ref{qsvt}). Solid lines corresponds to the CI-MRL calculations while the dashed lines are those performed in the CI-RL truncation. For each case, the vertical lines in the time-like region denote the indicate of the mass pole $Q^2=-m_T^2$.\label{qtv}}
\end{figure}

\section{Pion scalar, vector and tensor form factors}\label{sec:ff}
\begin{figure}[h!]
\includegraphics[width=8.6cm]{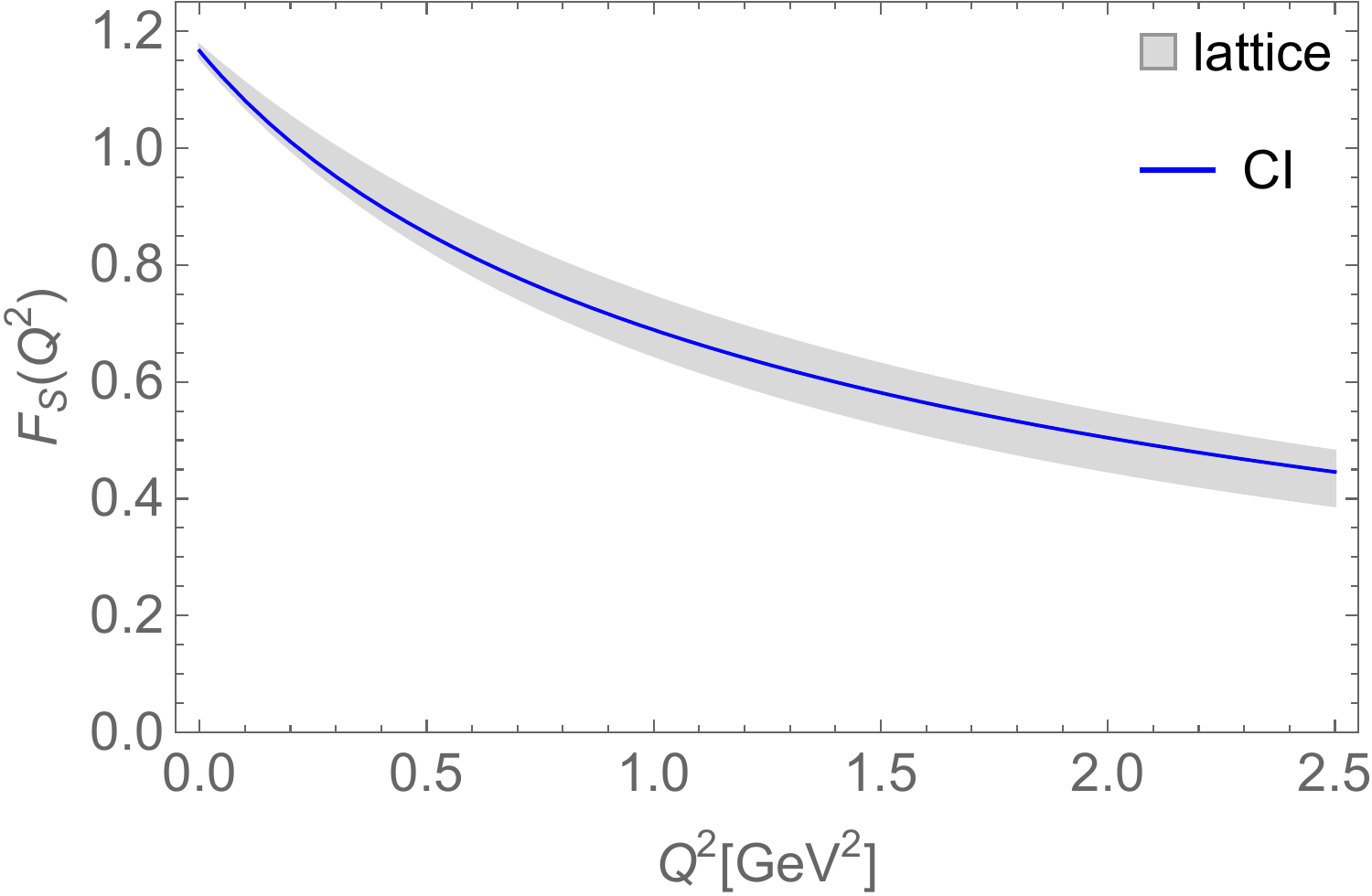}
\caption{Pion scalar form factor $F_{S}^{\pi}(Q^2)$. The solid line is our computed CI-MRL result, using the vertex dressing from Eq.~(\ref{fsb}). The band corresponds to the lattice QCD calculation from Ref.~\cite{Alexandrou:2021ztx}.\label{ffs}}
\end{figure}

\begin{figure}[h!]
\includegraphics[width=8.6cm]{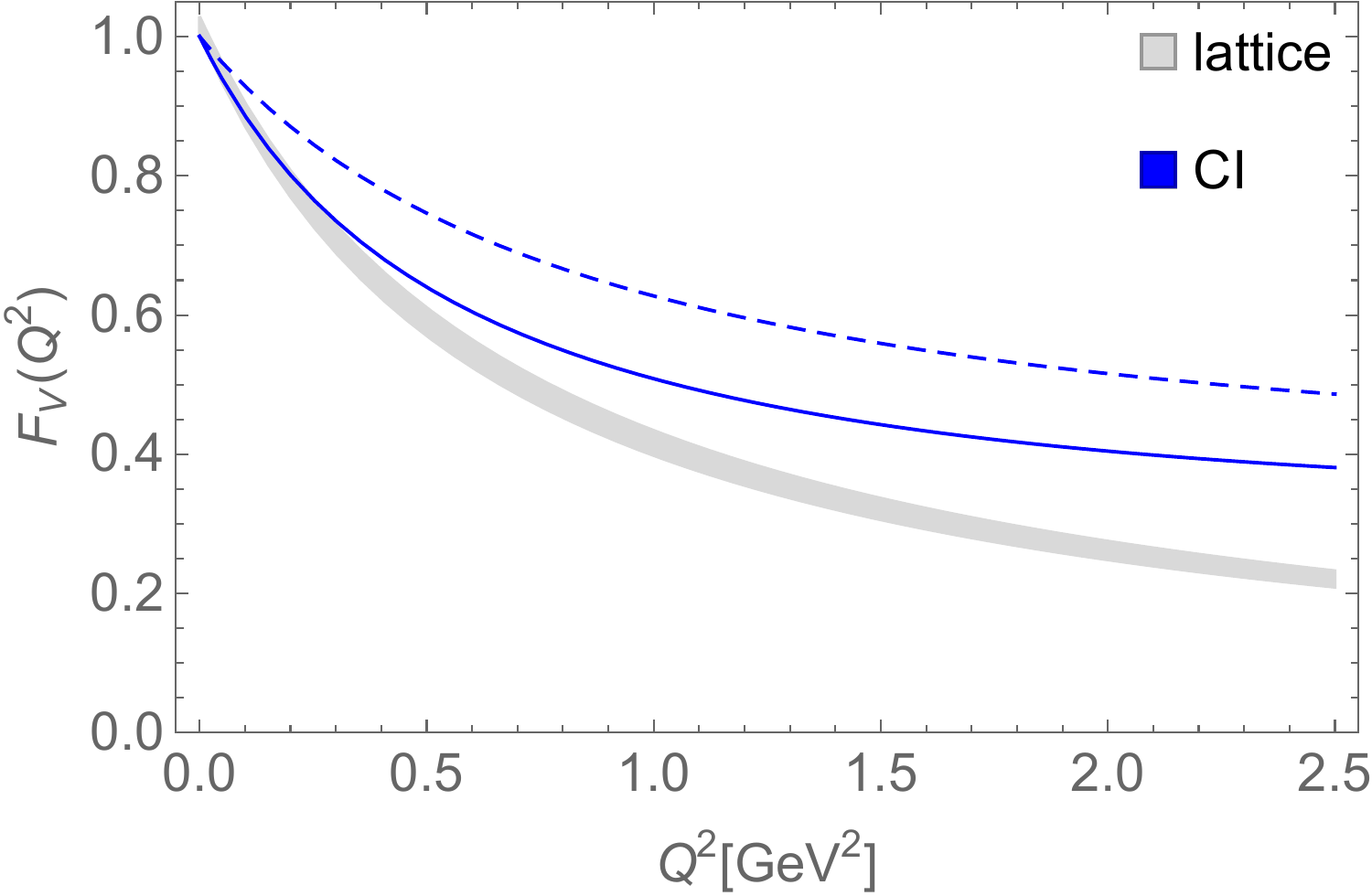}
\caption{Pion vector form factor $F_{V}^{\pi}(Q^2)$. The solid line corresponds to our result using the CI-MRL truncation, while  the dotted line is the analogous for the CI-RL case. The band corresponds to the lattice QCD calculation from Ref.~\cite{Alexandrou:2021ztx}.\label{ffv}}
\end{figure}

\begin{figure}[h!]
\includegraphics[width=8.6cm]{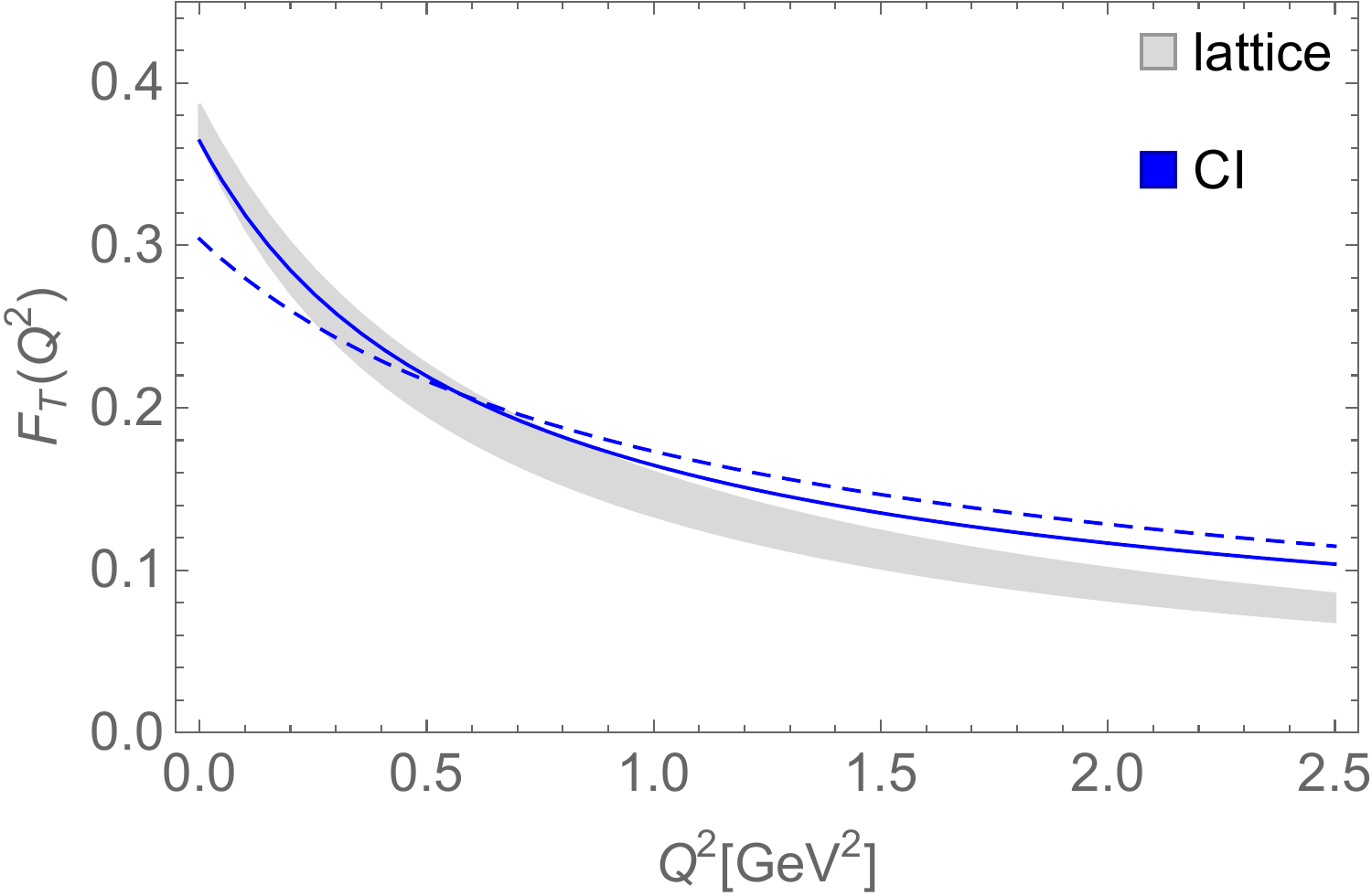}
\caption{Pion tensor form factor $F_{T}^{\pi}(Q^2)$. The solid line corresponds to our result using the CI-MRL truncation, while the dotted line is the analogous for the CI-RL case. The band corresponds to the lattice QCD calculation from Ref.~\cite{Alexandrou:2021ztx}.\label{fft}}
\end{figure}

With the quark propagators, pion Bethe-Salpeter amplitude and corresponding interaction vertices at hand, we are now in position to compute the scalar, vector and tensor FFs in the impulse approximation (IA)~\cite{Maris:2000sk}:
\begin{eqnarray}
    \label{eq:IA}
    T^{\#}(K,Q)F_{\#}^{\pi}(Q^2) &&= 2 \text{tr}_{CD} \int_q \left\{i\Gamma_\pi(-p_f)S(q+p_f)\right.\nonumber\\ 
&&\left.\times i\Gamma^{\#}(Q)S(q+p_i)i\Gamma_\pi(p_i)S(q)\right\}\,,
\end{eqnarray}
where $\text{tr}_{CD}$ indicates trace over color and Dirac indices. The label $\#$ refers to scalar, vector and tensor cases such that, naturally, $\Gamma^{\#}$ are the interaction vertices from Eq.~\eqref{qsvt}, while $F_{\#}$ correspond to the FFs dressing the the matrix elements $T^{\#}(K,Q)$:
\begin{eqnarray}
&&T^{S}(K,Q)=i \mathbb{I}\,,\\
&&T^{V}_\mu(K,Q)=2K_{\mu}\,,\\
&&T^{T}_{\mu\nu}(K,Q)=-\frac{K_{\mu}Q_{\nu}-K_{\nu}Q_{\mu}}{m_{\pi}}\,.
\label{eq:matelem}
\end{eqnarray}
The kinematics is defined as follows: $Q=p_f-p_i$ is the momentum of the incoming gauge boson and $p_{i,f}=K\mp\frac{Q}{2}$ are the incoming and outgoing pion momentum, respectively; on-shell conditions entail $p_{i,f}^2-m_\pi^2$, such that $K\cdot Q=0$ and $K^2=-m_\pi^2-\frac{Q^2}{4}$. The key steps in calculation are presented in Appendix~\ref{App::keysteps}. The computed form factors are displayed in Figs.~\ref{ffs}-\ref{fft}.

Our final result for the pion sFF agrees fairly well with that from the lattice calculation in Ref.~\cite{Alexandrou:2021ztx}. A crucial piece for such outcome is the monopole Ansatz for the quark-scalar dressing function, $\bar{f}_S(Q^2)$, given in Eq.~\eqref{fsb}. If one were to take $f_S(Q^2)$ instead, the actual solution of the corresponding BSE, the very simple structure of $\Gamma_S$ permitted by the CI model would produce a notoriously harder sFF. This of course could be anticipated from Fig.~\ref{qsv}, highlighting the need of as artificial improvement of the quark-scalar vertex within the CI framework.

The vector form factor, namely the electromagnetic form factor, is depicted in Fig.~\ref{ffv}. Compared to the CI-RL result, the one obtained from the CI-MRL truncation is suppressed and much closer to the lattice QCD result. The suppressing effect comes, mainly, from the AMM term in the quark-photon vertex, but also due to the smaller computed value of $m_\rho$, which moves the vector meson pole closer to the space-like axis and, consequently, produces a larger charge radius. As the momentum transfer increases, the flaws of the CI are exposed and the difference between the CI calculations and lattice QCD becomes larger. In general, one would expect harder form factors to be produced by the CI interaction model~\cite{Wilson:2011aa,Raya:2021pyr}, since the nature of the CI implies constant mass functions and BSAs, which in turn lead to much simpler interaction vertices. In particular for the pion, it is well known that a symmetry-preserving treatment of the CI incorporates a pseudovector component in the pion BSA, Eq.~\eqref{bsa}, which eventually produces a pion eFF~\cite{Gutierrez-Guerrero:2010waf}  with an asymptotic behavior in marked contradiction with the prescriptions of perturbative QCD~\cite{Lepage:1980fj}; a similar outcome occurs for the pion to two-photon transition form factors~\cite{Roberts:2010rn}. Leaving the hardness issues aside, it is still notorious that, unlike the CI-RL case, our CI-MRL approach produces a vector FF compatible with that obtained from lattice QCD in a low $Q^2$ region.

\begin{figure}[h!]
\includegraphics[width=8.6cm]{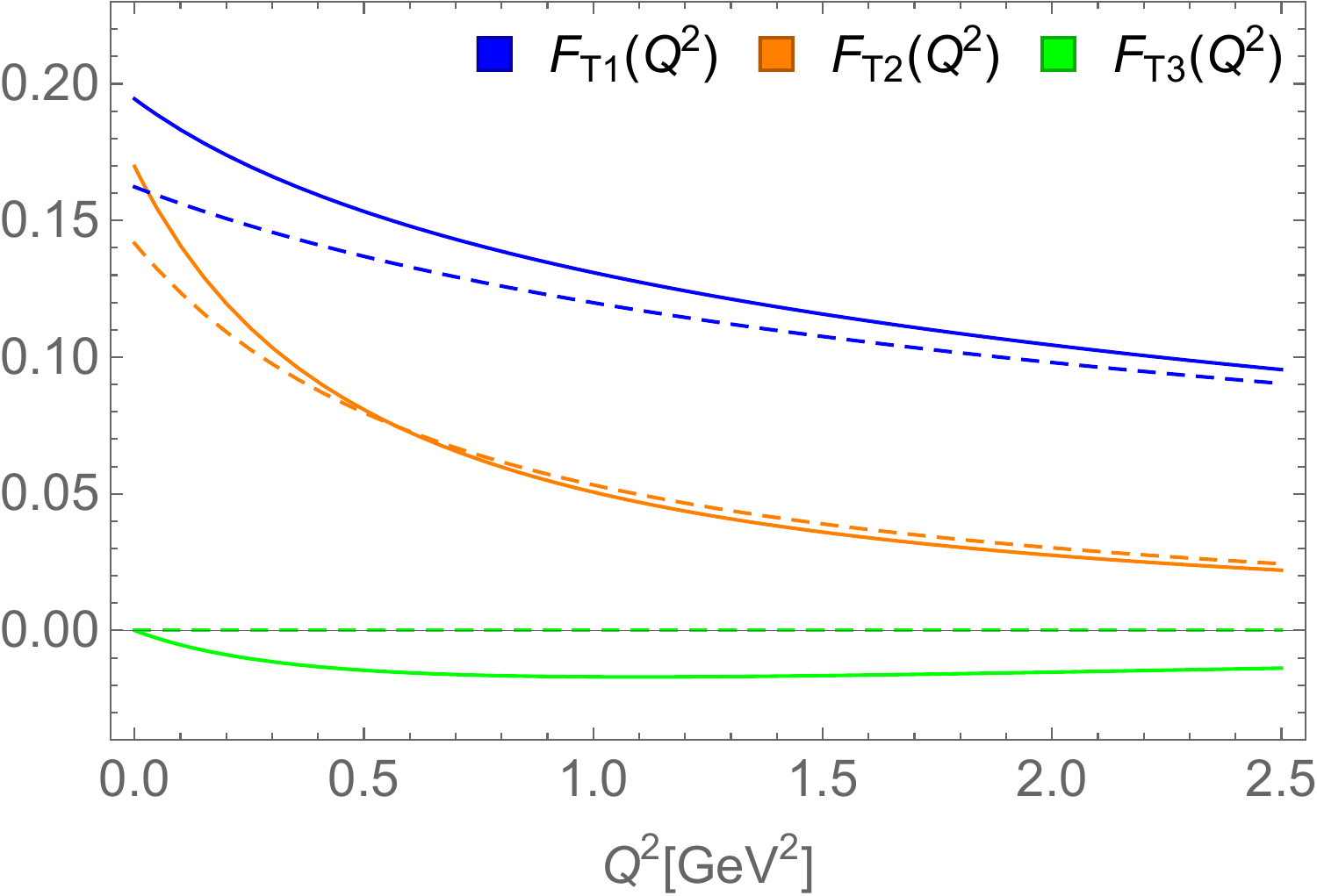}
\caption{The ingredients of the form factor from the 3 structures of the quark-tensor vertex. The solid line is computed using CI-MRL and the dashed line is computed with CI-RL.\label{tpart}}
\end{figure}

In the case of the tFF, the triangle diagram from Eq.~\eqref{eq:IA} is, in principle, insufficient to satisfy the WGTI; namely, charge conservation is not ensured by means of having $Q_\mu T^{\mu\nu}_T(K,Q) \neq 0$. This requires the IA to be completed with other diagrams which turn out to be proportional to $g_{\mu\nu}$; for self-consistency, $\Gamma_{\mu\nu}^T$ in Eq.~\eqref{qsvt} would also require such completion, ensuring the quark-tensor vertex satisfies a WGTI of its own~\cite{Brout:1966oea}. Notwithstanding, as discussed in Appendix A, the tFF is decoupled from its matrix element after contracting with $K_\mu Q_\nu$; therefore, due to the on-shell condition $K\cdot Q = 0$, any contribution coming from $g_{\mu\nu}$ vanishes. This makes the present approach sufficient for a consistent description of the tFF. Our results in this case are presented in Figs.~(\ref{fft}, \ref{tpart}). The former depicts our final outcomes in the CI-RL and CI-MRL truncations, as compared with those from Ref.~\cite{Alexandrou:2021ztx}; the latter dissects the individual contributions of the QT vertex pieces to the tFF. Focusing on Fig.~(\ref{fft}), it is clear that the CI-MRL calculation improves that obtained within CI-RL, producing a tFF in keen agreement with the lattice result over a rather generous domain of $Q^2$. Again, there an increasing discrepancy between the CI predictions and lattice simulation is expected as $Q^2$ becomes larger. To further scrutinize on the impacts of the NL term on the computed tFF, we separate the contributions of each piece of the quark-tensor vertex in Eq.~(\ref{qsvt}) to the tFF,
\begin{equation}
F_{T}^{\pi}(Q^2)=F_{T1}^{\pi}(Q^2)+F_{T2}^{\pi}(Q^2)+F_{T3}^{\pi}(Q^2).
\end{equation}
The individual contributions are depicted in Fig.~\ref{tpart}. Capitalizing on $F_{T1, T2}^\pi(Q^2)$, it is clear that the difference between the CI-RL and CI-MRL results is essentially due to the $Q^2=-m_T^2$ pole. For $F_{T2}^\pi(Q^2)$, the poles are shifted according to the computed value of $m_T$, while for $F_{T1}^\pi(Q^2)$, the mass pole is completely removed in the CI-RL case. As discussed in Sec.~\ref{sec:qsvt}, the presence of the NL term implies a decreasing in the value of $m_T$, letting the pole position move closer to $Q^2=0$ and, consequently, the form factors $F_{T1,T2}^{\pi}(Q^2)$ exhibit a steeper falloff in the low $Q^2$ space-like domain. Regarding $F_{T3}^\pi(Q^2)$, the third structure structure characterizing the quark-tensor vertex, $\frac{i^2}{M^2}\slashed{Q}\sigma_{\mu\nu}\slashed{Q}\sim Q^2$, only survives if the NL term appears the corresponding Bethe-Salpeter kernel. Furthermore, the projection operator that decouples the tFF from its matrix element produces a vanishing $F_{T3}^{\pi}(Q^2)$ at $Q^2=0$. Conversely, as $Q^2$ increases, the CI-MRL obtained form factor $F_{T}^{\pi}(Q^2)$ is suppressed mainly because of the negative contribution given by $F_{T3}^{\pi}(Q^2)$. The destructive interference of $F_{T1,T2}(Q^2)$ with $F_{T3}(Q^2)$ then produces softer behavior at large $Q^2$.


\begin{figure}[h!]
\includegraphics[width=8.6cm]{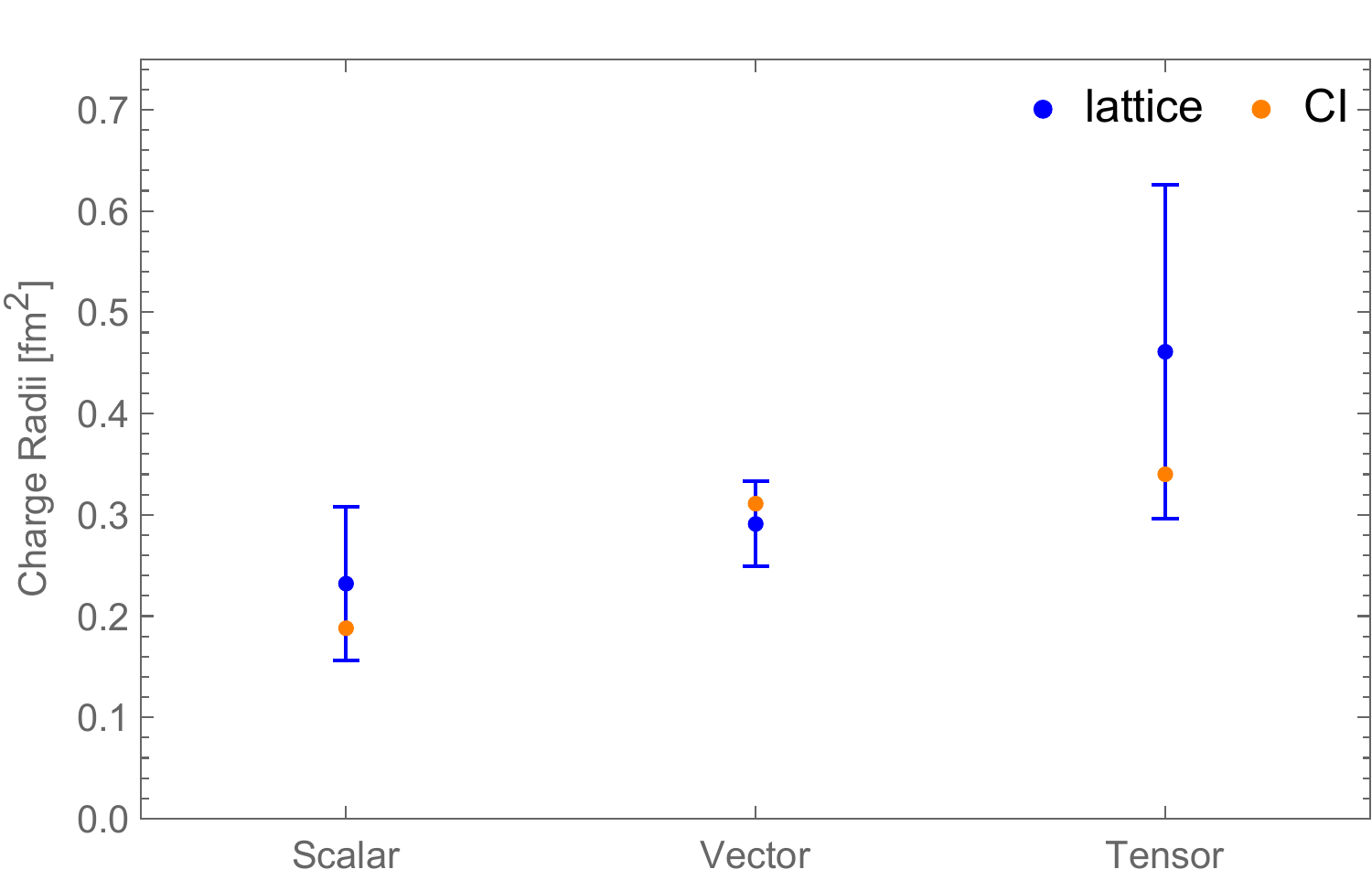}
\caption{Scalar, vector and tensor charge radii, as defined in Eq.~\eqref{eq:radiidef}. The orange points corresponds to values obtained within the CI-MRL scheme, while the blue data points are those from lattice QCD~\cite{Alexandrou:2021ztx}.\label{radii}}
\end{figure}

For a final comparisson, we compute the corresponding charge radii, defined as
\begin{eqnarray}
\label{eq:radiidef}
\<r^2\>_{\#}=-\left.\frac{6}{F_{\#}(0)}\frac{\partial F_{\#}(Q^2)}{\partial Q^2}\right|_{Q^2=0},
\end{eqnarray}
and producing the CI-MRL inferred values of:
\begin{eqnarray}
&&{r_{S}^{\pi}}^2=(0.434\,\text{fm})^2\,,\nonumber\\
&&{r_{V}^{\pi}}^2=(0.558\,\text{fm})^2\,,\nonumber\\
&&{r_{T}^{\pi}}^2=(0.583\,\text{fm})^2\,.\label{eq:radii}
\end{eqnarray}
A first thing to notice is the pattern $r_S \textless r_V \lesssim r_T$, which is indeed is the same followed by the inverse of the masses, \emph{i.e.} $1/m_\sigma \textless 1/m_\rho = 1/m_T$. The comparison between the CI-MRL and lattice results are shown in Fig.~\ref{radii}. Since the CI-MRL exhibits comparable values and slopes of the vector and tensor form factors at $Q^2=0$, the charge radii are in agreement with lattice QCD simulations. In comparisson with traditional CI calculations (for instance, Ref.~\cite{Roberts:2011wy}), the value $r_{V}^\pi = 0.558\,\text{fm}$ lies closer to the experimental one~\cite{Zyla:2020zbs}, $r_{V}^{exp} = 0.659(4)$ fm, and to that from the analysis of Ref·\cite{Cui:2021aee}, $r_{V}^{spm}=0.640(7)$ fm.  This outcome is a consequence of the richer structure of the quark-photon vertex: roughly, the contribution from the $f_{V2}$ dressing (the one containing the vector meson pole) enhances by $50~\%$ that comming from $\gamma_\mu$ alone, while $f_{V3}$, the AMM piece, further enhances such value by $25~\%$. A similar reasoning could be applied to $r_T^\pi$ as well, where the $\sigma_{\mu\nu}$ term alone (in the QT vertex) produces a charge radius about $25~\%$ smaller than the one reported in Eq.~\eqref{eq:radii}.

\section{summary}\label{sec:sum}
In this article, the pion scalar, vector and tensor FFs, as well as the corresponding charge radii, are investigated within a contact interaction model of QCD. Following a previous study in the vector meson sector~\cite{Xing:2021dwe}, the Bethe-Salpeter kernel describing the scalar, vector and tensor channels, is complemented with new structures in addition to those coming from the rainbow-ladder truncation alone. Despite this, and just as the CI-RL does, the CI-MRL truncation is consistent with the vector and axial-vector WGTIs. In implementing this truncation, the structure of the quark-vector (quark-photon) and quark-tensor vertices is enriched, with respect to its CI-RL counterpart, and we have seen that this leads to a positive impact on the FFs under study. Comparing the produced FFs with thos obtained from recent lattice QCD~\cite{Alexandrou:2021ztx}, we can conclude the following:

Even though the momentum independent nature of the CI model produces, typically, harder form factors, the CI-MRL truncation brings some improvements in the $Q^2$ dependence of such. When computed under the CI-MRL scheme, the scalar, vector and tensor from factors display a better compatibility with the lattice results at space-like momenta; much better than in the CI-RL case. This is a consequence of the profile that the interaction vertices acquire under this truncation. For instance, the quark-vector (quark-photon) vertex contains an AMM term in addition to the transverse piece containing the vector meson pole. 

On the other hand, the QT vertex exhibits a non trivial profile in the three dressing functions characterizing it; two of them feature a mass pole in the timelike axis, which produces both a larger charge radius and a better behavior at small $Q^2$; a third one, $f_{T3}$, which is non-zero only in the CI-MRL case, modulates the behavior at large $Q^2$ in such a way that one obtains a softer tFF, in better agreement with the lattice QCD result. This would imply that for the tFF, the CI-MRL leads to improvements at both small and large $Q^2$. For the scalar case, however, the QS vertex dressing function one obtains from its Bethe-Salpeter equation is not sufficiently damped. This would produce a sFF way too hard, making it necessary to introduce a sensible Ansatz for the vertex dressing function, in order to get a proper description. This crucial drawback is well understood from the very simple structure that the quark-scalar vertex acquires in the CI model, which prevails in both CI-RL and CI-MRL truncations. If computed within the CI-MRL model, all charge radii computed herein are  plainly compatible with those from lattice QCD. Contrary to the CI-RL case, the mass of the vector meson can be placed at its physical value, so that the pole impacts the small $Q^2$ region to a greater extent and, thus enlarging the values of the charge radii. The AMM term appearing in the quark-photon vertex, also plays a positive role in getting a much sensible value for the radii. 

Therefore, all virtues of the computed FFs and corresponding charge radii are attributed to the extended structure of the symmetry-preserving CI-MRL truncation, which is also known to significantly improve the description of the static properties and electromagnetic FFs of the $\rho$ meson~\cite{Xing:2021dwe}. Further extensions concerning the axial-vector meson are currently being investigated. 

\section{Acknowledgements}
KR acknowledges valuable discussions with J. Rodríguez Quintero. Work supported by National Natural Science Foundation of China (grant no.\,12135007).
\appendix\begin{widetext}
\section{Quark-tensor vertex and tensor form factor}\label{App::keysteps}
This appendix shows some key steps in the calculation of the dressed quark-tensor vertex and corresponding pion tensor form factor. The steps described herein are quite general and applicable for the rest of the cases. 

Let us start from the dressed quark-tensor vertex, which satisfies the following inhomogeneous BSE:
\begin{equation}\label{a1}
    \begin{aligned}
        \Gamma_{\mu\nu}^{T}(Q)=\sigma_{\mu\nu}&-\frac{4}{3m_G^2}\int_q \gamma_\alpha S(q)\Gamma_{\mu\nu}^{T}(Q)S(q-Q)\gamma_\alpha+\frac{4\xi}{3m_G^2}\int_q \tilde{\Gamma}_n S(q)\Gamma_{\mu\nu}^{T}(Q)S(q-Q)\tilde{\Gamma}_n
    \end{aligned}
\end{equation}
where $\tilde{\Gamma}_n=\{\mathbb{I},\gamma_5,\frac{i}{\sqrt{6}}\sigma_{\mu\nu}\}$, and
\begin{equation}\label{eq:a2}
	\Gamma_{\mu\nu}^T(Q)=f_{T1}(Q^2)\sigma_{\mu\nu}+f_{T2}(Q^2)\frac{i}{M}(\slashed{Q}\sigma_{\mu\nu}-\sigma_{\mu\nu}\slashed{Q})+f_{T3}(Q^2)\frac{i^2}{M^2}\slashed{Q}\sigma_{\mu\nu}\slashed{Q}\;.
\end{equation}
Using the above equation, the main idea is to conveniently recast Eq.~\eqref{a1} as follows:
\begin{equation}
	\begin{pmatrix}
		f_{T1}(Q^2)\\
		f_{T2}(Q^2)\\
		f_{T3}(Q^2)
	\end{pmatrix}
	=\begin{pmatrix}
		a_{11}&a_{12}&a_{13}\\
		a_{21}&a_{22}&a_{23}\\
		a_{31}&a_{32}&a_{33}
	\end{pmatrix}
	\begin{pmatrix}
		f_{T1}(Q^2)\\
		f_{T2}(Q^2)\\
		f_{T3}(Q^2)
	\end{pmatrix}+
	\begin{pmatrix}
		1\\
		0\\
		0
	\end{pmatrix}
\end{equation}
Because the three basis elements of $\Gamma_{\mu\nu}^{T}(Q)$ are Dirac-trace orthogonal, we can easily calculate the matrix entries $a_{i,j}$. In fact, it is sufficient to multiply Eq.~\eqref{a1} by each element from the basis in Eq.~\eqref{eq:a2}. For instance, one has:
\begin{equation}
	\begin{aligned}
		a_{22}&={\text{tr}_{CD}}\left\{\frac{\slashed{Q}\sigma_{\mu\nu}-\sigma_{\mu\nu}\slashed{Q}}{72m_G^2Q^2}\left[\int_q\gamma_\alpha S(q)(\slashed{Q}\sigma_{\mu\nu}-\sigma_{\mu\nu}\slashed{Q})S(q-Q)\gamma_\alpha-\int_q\xi\tilde{\Gamma}_n S(q)(\slashed{Q}\sigma_{\mu\nu}-\sigma_{\mu\nu}\slashed{Q})S(q-Q)\tilde{\Gamma}_n\right]\right\}\\
		&=\frac{4}{3m_G^2}\int_{0}^{1}d\alpha\int_q\frac{q^2-2\alpha(1-\alpha)Q^2+2M^2}{(q^2+\alpha(1-\alpha)Q^2+M^2)^2}\;,
	\end{aligned}
\end{equation}
where the second line has been obtained after a combination of Feynman parametrization and changes of variables. Then, one can choose whether to use the axial-vector WGTI~\cite{Roberts:2010rn}, in order to get rid of the integral involving a $q^2$ in the numerator. As a matter of fact, it is not necessary for the 4-momentum integrals concerning the quark-scalar vertex, but it is quite convenient in the vector and tensor cases. In particular, still for $a_{22}$, one gets:
\begin{equation}
	a_{22}=-Q^2 \tilde{I}(Q^2)\;,
\end{equation}
where the following definitions have been employed:
\begin{eqnarray}
	\tilde{I}(Q^2)&=&\frac{1}{3\pi^2m_G^2}\int_{0}^{1}\alpha(1-\alpha)\bar{\mathcal{C}}_{1}^{\text{iu}}(\omega(\alpha,Q^2))d\alpha\;,
\\
	\bar{\mathcal{C}}_{1}^{\text{iu}}(\omega)&=&-\frac{d}{d\omega}\mathcal{C}^{iu}(\omega)\;,\\
	\omega(\alpha,Q^2)&=&\alpha(1-\alpha)Q^2+M^2\;.
\end{eqnarray}
Proceeding analogously with the rest of the coeficcients $a_{ij}$, one obtains the following expressions for the quark-tensor vertex dressing functions:
\begin{equation}
	\begin{aligned}
		&f_{T1}(Q^2)=\frac{-3(\xi M^2Q^2I(Q^2)^2-4\xi M^2Q^2I(Q^2)\tilde{I}(Q^2)-4\xi M^2I(Q^2)+6Q^2I(Q^2)+6)}{2(2\xi M^2I(Q^2)+2\xi Q^2\tilde{I}(Q^2)-3)[(1+Q^2\tilde{I}(Q^2))(-2\xi M^2I(Q^2)+2\xi Q^2\tilde{I}(Q^2)+3)+\xi M^2Q^2I(Q^2)^2]}\,,\\
		&f_{T2}(Q^2)=\frac{-3M^2I(Q^2)}{4[(1+Q^2\tilde{I}(Q^2))(-2\xi M^2I(Q^2)+2\xi Q^2\tilde{I}(Q^2)+3)+\xi M^2Q^2I(Q^2)^2]}\,,\\
		&f_{T3}(Q^2)=\frac{3\xi M^2(M^2I(Q^2)^2+4Q^2\tilde{I}(Q^2)^2+4\tilde{I}(Q^2))}{2(2\xi M^2I(Q^2)+2\xi Q^2\tilde{I}(Q^2)-3)[(1+Q^2\tilde{I}(Q^2))(-2\xi M^2I(Q^2)+2\xi Q^2\tilde{I}(Q^2)+3)+\xi M^2Q^2I(Q^2)^2]}\,,
	\end{aligned}
\end{equation}
where
\begin{eqnarray}
    I(Q^2)&=&\frac{1}{3\pi^2m_G^2}\int_{0}^{1}\bar{\mathcal{C}}_{1}^{\text{iu}}(\omega(\alpha,Q^2))d\alpha\;.
\end{eqnarray}
To get the tFF from its matrix element, we now multiply both left and right hand sides of Eq.~\eqref{eq:IA} by $K_{\mu}Q_{\nu}$, such that
\begin{equation}
	\begin{aligned}
	    -\frac{K^2Q^2}{m_{\pi}}F_{V}^{\pi}(Q^2)&=2 \text{tr}_{CD}\int_q \bigg\{i\Gamma_\pi(-p_f)S(q+p_f)i\left[f_{T1}(Q^2)i\slashed{K}\slashed{Q}+f_{T2}(Q^2)\frac{2Q^2\slashed{K}}{M}+f_{T3}(Q^2)\frac{iQ^2\slashed{K}\slashed{Q}}{M^2}\right]\\
	    &\times S(q+p_i)i\Gamma_\pi(p_i)S(q)\bigg\}
	\end{aligned}
\end{equation}
Given the structure of the pion BSA, Eq.~\eqref{bsa}, the tensor form factor can be splitted in three parts: one proportional to $(E_{\pi}^c)^2$, an analogous for $(F_{\pi}^c)^2$, and a third one containing the crossed term $E_{\pi}^cF_{\pi}^c$. In this case, the one with $(F_{\pi}^c)^2$ can be evaluated directly, while the other two requires some algebraic manipulations to ensure the axial WGTI and translational invariance in the CI~\cite{Roberts:2010rn}. For instance, the following expression accompanies the $(E_{\pi}^c)^2$ term:
\begin{equation}
	\begin{aligned}
		&\frac{8N_cm_{\pi}}{M(m_{\pi}^2+Q^2/4)}\int_q\frac{(m_{\pi}^2+Q^2/4)(M^2f_{T1}(Q^2)-4(q^2+M^2)f_{T2}(Q^2)+Q^2f_{T3}(Q^2))+8(K\cdot q)(q^2+M^2-m_{\pi}^2)f_{T2}(Q^2)}{((q+K+Q/2)^2+M^2)((q+K-Q/2)^2+M^2)(q^2+M^2)}\\
		&=\frac{8N_cm_{\pi}}{M(m_{\pi}^2+Q^2/4)}\int_q\left[\frac{-4(m_{\pi}^2+Q^2/4-2(K\cdot q))f_{T2}(Q^2)}{((q+K+Q/2)^2+M^2)((q+K-Q/2)^2+M^2)}\right.\\
		&\left.+\frac{(m_{\pi}^2+Q^2/4)(M^2f_{T1}(Q^2)+Q^2f_{T3}(Q^2))-8(K\cdot q)m_{\pi}^2f_{T2}(Q^2)}{((q+K+Q/2)^2+M^2)((q+K-Q/2)^2+M^2)(q^2+M^2)}\right]
	\end{aligned}
\end{equation}
As with the QT vertex, from a combination of Feynman parameterization and changes of variables, it is possible to evaluate each addend separately. For the crossed term of $E_{\pi}^c F_{\pi}^c$, we follow a similar approach. Finally, one arrives at:
\begin{equation}
	\begin{aligned}
		F_{T}^{\pi}(Q^2)&=\frac{N_cm_{\pi}}{2\pi^2M^3}\left\{E_{\pi}^c[M^2F_{\pi}^cf_{T1}(Q^2)+2M^2(2F_{\pi}^c-E_{\pi}^c)f_{T2}(Q^2)+Q^2f_{T3}(Q^2)]\int_{0}^{1}d\alpha \bar{\mathcal{C}}_{1}^{\text{iu}}(\omega(\alpha,Q^2))\right.\\
		&-2F_{\pi}^c\int_{0}^{1}du_1\int_{0}^{1-u_1}du_2[M^2(E_{\pi}^c+F_{\pi}^c)f_{T1}(Q^2)+((3u_1+3u_2-2)m_{\pi}^2+(u_1+u_2)Q^2)F_{\pi}^cf_{T2}(Q^2)\\
		&+Q^2(E_{\pi}^c+F_{\pi}^c)f_{T3}(Q^2)]\bar{\mathcal{C}}_{1}^{\text{iu}}(\omega'(u_1,u_2,Q^2))\\
		&+2\int_{0}^{1}du_1\int_{0}^{1-u_1}du_2[(M^2E_{\pi}^{c2}+(2M^2+(u_1+u_2)m_{\pi}^2)F_{\pi}^{c2}\\
		&+(-3M^2+(3u_1^2+2u_1(3u_2-1)+u_2(3u_2-2))m_{\pi}^2+3u_1u_2Q^2)E_{\pi}^cF_{\pi}^c)(M^2f_{T1}(Q^2)+Q^2f_{T3}(Q^2))\\
		&+(-2(u_1+u_2)m_{\pi}^2M^2E_{\pi}^{c2}+2M^2(2m_{\pi}^2-(u_1+u_2)Q^2)E_{\pi}^cF_{\pi}^c\\
		&+(u_1u_2(u_1+u_2)Q^4+(u_1+u_2-1)(u_1+u_2)^2m_{\pi}^4+(5u_1+5u_2-6)m_{\pi}^2M^2+3(u_1+u_2)Q^2M^2\\
		&+(-u_1^3+u_1^2(6u_2+1)+2u_1u_2(3u_2-2)-u_2^2(u_2-1))m_{\pi}^2Q^2)F_{\pi}^{c2})f_{T2}(Q^2)]\bar{\mathcal{C}}_{2}^{\text{iu}}(\omega'(u_1,u_2,Q^2))\bigg\}\,
	\end{aligned}
\end{equation}
where
\begin{eqnarray}
	\bar{\mathcal{C}}_{2}^{\text{iu}}(\omega)&=&\frac{1}{2}\frac{d^2}{dw^2}\mathcal{C}^{iu}(\omega)\\
	\omega'(u_1,u_2,Q^2)&=&M^2+m_{\pi}^2(u_1+u_2)(u_1+u_2-1)+u_1u_2Q^2
\end{eqnarray}
\clearpage
\end{widetext}
\clearpage
\newpage
\bibliography{main}

\end{document}